# Position-Sensitive Detector with Depth-of-Interaction Determination for Small Animal PET[*)]

A. Fedorov[1], A. Annenkov[2], A. Kholmetsky[1], M. Korzhik[1], P. Lecoq[3], A. Lobko[1], O. Missevitch[1], A. Tkatchov[2]

1-Institute for Nuclear Problems, Minsk, Belarus
2-Bogoroditsk Technical Chemical Plant, Bogoroditsk, Tula Region, Russia
3-CERN, Geneva, Switzerland

*Abstract*

Crystal arrays made of $Lu_2SiO_5$:Ce (LSO) and $LuAlO_3$:Ce (LuAP) scintillation crystal pixels with 2×2×10 mm³ dimensions were manufactured for evaluation of detector with depth-of-interaction (DOI) determination capability intended for small animal positron emission tomograph (PET). Two designs of dual-layer scintillation detector based on the position-sensitive multi-anode photomultiplier tube, where acted scintillation layer is either identified by pulse shape discrimination or directly decoded from photomultiplier anode signals distribution, were tested and compared. Position-sensitive LSO/LuAP phoswich detector with DOI determination capability for small animal PET based on crystal arrays consist of 8×8 pixels and HAMAMATSU R5900-00-M64 position-sensitive multi-anode photomultiplier tube (PSPMT) was developed and evaluated. Time resolution was found to be not worse than 1.0 ns FWHM for both layers, and spatial resolution mean value was 1.5 mm FWHM for the center of field-of-view (FOV). Design of three-layer scintillation detector with DOI for small animal PET was proposed.

## I. INTRODUCTION

For the development of new human radiopharmaceuticals and drugs and for their operation clarification, studies on laboratory animals can be undertaken before human ones. PET studies of the tracer biodistribution and pharmacokinetics will require less laboratory animals in comparison with other techniques which is significant aspect connected with limitations on animal experiments. Human PET scanners can be used for this purpose, however this is unpractical due to their expensiveness. Moreover, small animal studies on clinical PET scanners may be forbidden both in EU and USA.

Thus, the need for dedicated PET scanners designed for experimental small animal (mouse, rat, rabbit) studies was recognized [1]. Here, a device with moderate price and high spatial resolution over limited field of view can be used in pharmacology for new drug development and their mechanism studies, for evaluation of results of the tumor therapy and for neuroscience studies.

Required level of the spatial resolution for a small animal PET would be in range from 1.0 mm to 1.5 mm FWHM [2]. As far as such requirements are rather close to theoretical limits [3], use of Anger Logic which brings additional uncertainty in hit point determination is undesirable. New approaches to construct a high-resolution PET scanner may consist in use of new scintillators with short decay time and high stopping power read out by position-sensitive photo detectors.

On the other hand, sensitivity of commercially offered by Concorde variant of MicroPET system with its 5400 cps/MBq, obtained with 10 mm long LSO crystals, seems to be minimum acceptable for practical use of the device. In consequence of this, DOI determination is the most efficient way to increase total thickness and sensitivity of the detector without degradation of its spatial resolution. From our point of view, the most convenient solution is the use of phoswich detector based on at least two layers of scintillators.

## II. DETECTOR DESIGN AND PERFOMANCE

Although technology of LuAP [4] growth is still under development, reached by now value of LuAP light yield allows to start feasibility study of PET detector on its base with photomultiplier tube readout. Due to significant difference in LuAP and LSO decay times (see Table 1) development of fast and high-sensitive LSO/LuAP phoswich detector seems to be the most promising application of LuAP in positron emission tomography.

Table 1
Comparison of some properties of LSO, LuAP and BGO scintillators

| Material | Density $\rho$, g/cm³ | Emission maximum $\lambda$, nm | Decay time $\tau$, ns | Photo-electric absorption coefficient @ 511 keV 1/cm |
|---|---|---|---|---|
| LSO | 7.4 | 420 | 40 | 0.30 |
| LuAP | 8.34 | 380 | 11 (60%) | 0.31 |
| BGO | 7.13 | 480 | 300 | 0.41 |

For evaluation of small animal PET phoswich detector with DOI determination capability, two arrays made of LSO and LuAP scintillation crystal 8×8 pixels with 2×2×10

[*)] This work is being carried out in the frame of the Crystal Clear Collaboration at CERN and is supported by ISTC.

mm$^3$ dimensions have been manufactured. These arrays were optically coupled together and mounted on HAMAMATSU R5900-00-M64 photomultiplier tube as is shown in Fig.1.

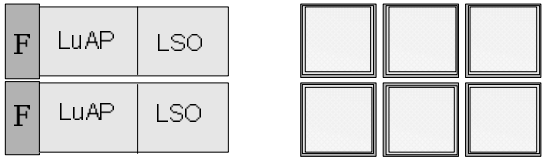

Figure 1: Detector design. *F* stands for one of 64 elements of the PMT anode pattern.

Time resolution of the detector was measured individually for upper and lower detector layers using collimated $^{22}$Na source. The experimental setup technically is the conventional start-stop time spectrometer consisting of constant fraction discriminators (CFD), delay lines, CAMAC time-to-amplitude converter (TAC), ADC and BaF$_2$ start detector. The time resolution was found to be not worse than 1.0 ns FWHM for both layers that allows to use 4-5 ns coincidence gate in PET data acquisition system.

The experimental setup used for evaluation of the small animal PET detector spatial resolution is based on NIM and CAMAC modules, Fig.2 below. It includes $^{22}$Na source with 0.5 mm collimator, reference detector made of 2×2×10 mm$^3$ LSO crystal and ADC card, gated by signals from fast coincidence unit (10 ns gate) and pulse shape discriminator (PSD). For the line spread function measurements the detector block was placed at 12 cm distance from the reference detector, and $^{22}$Na source with collimator was moved across the reference detector axis. Line spread functions were measured for both layers under angles between detector and collimator axis α=0-30°. For α=0° (center of FOV) spatial resolution mean value was found to be 1.5 mm FWHM.

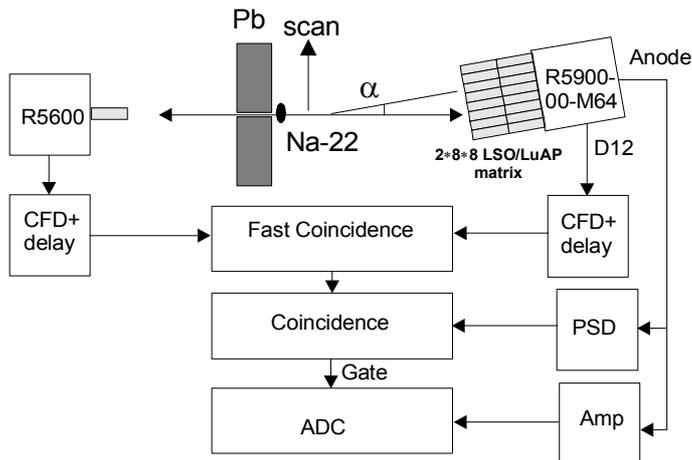

Figure 2: Experimental test setup.

Finally, an alternative design of the phoswich detector was tested using 10 pixel LSO array, see Fig.3. Here hit layer is uniquely decoded from photomultiplier anode signals distribution – scintillation light from the upper layer hits two adjacent PMT anode pixels, while scintillation in the lower layer hits single pixel. Evaluation study had shown operability of this design even in spite of significant light loss from the upper layer. Total charge collected from two adjacent PMT anodes is about 40% of charge produced by hit scintillation cell from the lower layer.

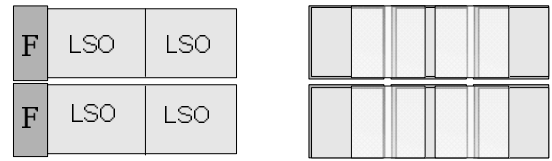

Figure 3: DOI detector with direct layer determination.

## III. CONCLUSIONS

The developed LSO/LuAP detector with PSPMT readout is the good approach for the development of small animal PET with DOI determination capability.

Improvement of LuAP light yield up to 12,000 – 15,000 ph/MeV will allow either to apply LuAP in detectors with light sharing (i.e. in one shown in Fig. 3) or to use avalanche photo diode (APD) readout.

Combination of schemes with pulse shape discrimination and light sharing will allow creating of multi-layer detectors with accurate DOI determination. Possible design of such detector is shown in Fig. 4.

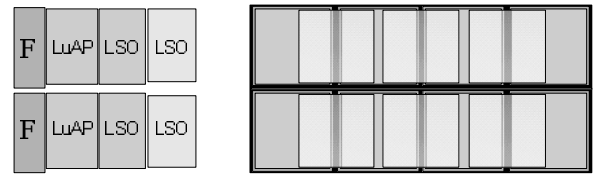

Figure 4: Three-layer DOI detector.